\documentstyle[twoside,fleqn,espcrc2,epsf]{article}

\newcommand{\latt}[2]{\mbox{${#1}^3 \times {#2}$}}

\newcommand{\Fig}[1]{\mbox{Fig.\ \ref{#1}}}
\newcommand{\Tab}[1]{\mbox{Table~\ref{#1}}}

\newcommand{\ie}{{\em i.e.\/}}

\newcommand{\etal}{{\em et al.\/}}

\newcommand{\err}[2]{\raisebox{0.08em}{\scriptsize {$\;\begin{array}{@{}l@{}}
			  +\makebox[0.95em][r]{#1} \\[-0.12em]
			  -\makebox[0.95em][r]{#2}
			\end{array}$}}}

\newcommand{\er}[2]{\raisebox{0.08em}{\scriptsize {$\;\begin{array}{@{}l@{}}
			  +\makebox[0.55em][r]{#1} \\[-0.12em]
			  -\makebox[0.55em][r]{#2}
			\end{array}$}}}

\newcommand{\ewxy}[2]{\setlength{\epsfxsize}{#2}\epsfbox[20 30 580 570]{#1}}

\newcommand{\yoshie}{Yoshi\'{e}}

\newcommand{\PS}{P\hspace{-0.10em}S}
\newcommand{\kappacrit}{\kappa_{\rm crit}}

\title{Quenched Light Hadron Spectroscopy: Comparing the
Wilson and $O(a)$-Improved Fermion Actions}

\author{{\it UKQCD Collaboration}\\
	presented by Alan D.\ Simpson
	\vskip\baselineskip{
	Department of Physics, Univerity of Edinburgh, \\
	The King's Buildings, Mayfield Road, Edinburgh, Scotland}
}

\begin{document}

\begin{abstract}

We have studied the light hadron spectrum and decay constants for
quenched QCD at $\beta=6.2$ on a \latt{24}{48} lattice. We
compare the results obtained using a nearest-neighbour $O(a)$-improved
(``clover'') fermion action with those obtained using the standard
Wilson fermion action on the same gauge configurations.
For pseudoscalar meson masses in the range 330-800~MeV, we find no
significant difference between the results for the two actions.  The
scales obtained from the string tension and mesonic sector are
consistent, but higher than those derived from baryon masses.  The
ratio of the pseudoscalar decay constant to the vector meson mass is
roughly independent of quark mass as observed experimentally.

\end{abstract}

\maketitle


\section{INTRODUCTION}

\subsection{$O(a)$-Improved Fermion Action}

Amongst the many systematic errors involved in the numerical simulation
of QCD are those due to the non-zero lattice spacing, $a$. The standard
Wilson pure gauge action differs from the continuum action by terms of
$O(a^2)$, while the Wilson fermion action introduces a term of $O(a)$
in order to avoid the fermion doubling problem.  Sheikholeslami and
Wohlert~\cite{sheikholeslami} proposed a form of the fermion action
which was shown by Heatlie~\etal~\cite{heatlie} to remove terms of
$O(a)$ from matrix elements in perturbation theory.  The improvement
procedure described in~\cite{heatlie} involves rotating the fermion
fields and adding a local ``clover''~\cite{hadrons_lett} term to
the Wilson action, \ie,
\begin{equation} S_F^C = S_F^W -
i\frac{\kappa}{2}\sum_{x,\mu,\nu}\bar{q}(x)
			F_{\mu\nu}(x)\sigma_{\mu\nu} q(x)  ,
\end{equation}
where $F_{\mu\nu}$ is
a lattice definition of the field strength tensor.

In this paper, we present results obtained for the light hadron spectrum
on a \latt{24}{48} lattice at $\beta=6.2$ using both the standard Wilson
action and this $O(a)$-improved clover action.

\subsection{Computational Details}

The simulations were performed on the Meiko i860 Computing Surface
sited in Edinburgh.  This consists of 64 nodes and has a sustained
performance for QCD simulations of 1 to 1.5~Gigaflops.  The gauge
configurations were generated using Hybrid Overrelaxation~\cite{npb}.
The overrelaxation was performed on a sequence of 3 $SU(2)$ subgroups
of $SU(3)$.  As this is a micro-canonical update, we follow every 5
overrelaxed sweeps with a Cabibbo-Marinari heatbath~\cite{C_M} sweep to
preserve ergodicity.  We allow 16800 sweeps for equilibration and
separate configurations by 2400 sweeps.

The comparison of the results for the two fermion actions is performed
on the same 18 configurations.  For each action, we have propagators at
5 different values of the hopping parameter, $\kappa$, chosen to give
roughly similar pseudoscalar masses in the range $330\to800$~MeV.  As
reported in~\cite{hadrons_lett}, we found that the correlators
obtained using the clover action were somewhat noisier and so we have
smeared the clover propagators at the sink to improve the signal.  We
also present results for the clover action on 36 configurations with 3
$\kappa$ values and a local sink.

The most computationally expensive part of this study was calculating
the quark propagators.  We find that red-black preconditioning improves
the efficiency by a factor of 3 and that, for the range of pseudoscalar
masses investigated, Minimal Residual is typically twice as efficient
as Conjugate Gradient.  In agreement with the study reported by Hockney
in~\cite{hockney}, we see only a small gain  from over-relaxation for
MR at $\beta=6.2$.


\section{RESULTS}

\subsection{Fitting Procedure}

Limited statistics and the amount of supercomputer time used imply that
it is essential to perform a careful analysis of the data.  All our
propagator fits take into account correlations between timeslices.  We
require that the fitting region, $[t_{\rm min}, t_{\rm max}]$,
satisfies the condition that small perturbations of $t_{\rm min}$ and
$t_{\rm max}$ give the same mass within errors.  We find that the
intervals $[12, 16]$ for local sinks and $[9, 13]$ for smeared sinks
are satisfactory.
The correlated $\chi^2 / \mbox{{\rm dof}}$ varies between 0.3 and 4,
which indicates both that we are taking correlations into account and
that we are getting reasonable fits.
Unlike the QCDPAX collaboration~\cite{yoshie}, we find that increasing
$t_{\rm min}$ does not significantly change the masses obtained.
However, we are only able to increase $t_{\rm min}$ slightly before our
signal becomes noisy.

The errors are estimated using a bootstrap procedure. $N_{\rm cfgs}$
configurations are chosen from the ensemble of $N_{\rm cfgs}$
configurations, allowing repetition.  The same analysis is performed on
this bootstrap sample as on the original $N_{\rm cfgs}$ configurations.
We repeat this typically 1000 times and determine the errors such that
the middle 68\% of our bootstrap samples lie within the error bars.


\subsection{Hadron Masses}

Throughout the range of quark mass used, we find  that $m_{\PS}^2$ is
linear with $1/2\kappa$ in accordance with PCAC.  From correlated fits
on 18 configurations, we obtain the following values of $\kappacrit$.
\begin{eqnarray}
\kappacrit^W & = & 0.15328\er{7}{4}\quad\mbox{(Wilson)} \nonumber \\
\kappacrit^C & = & 0.14311\er{6}{3}\quad\mbox{(clover)}
\label{eq:kappa_c}
\end{eqnarray}

\begin{figure}[ht]
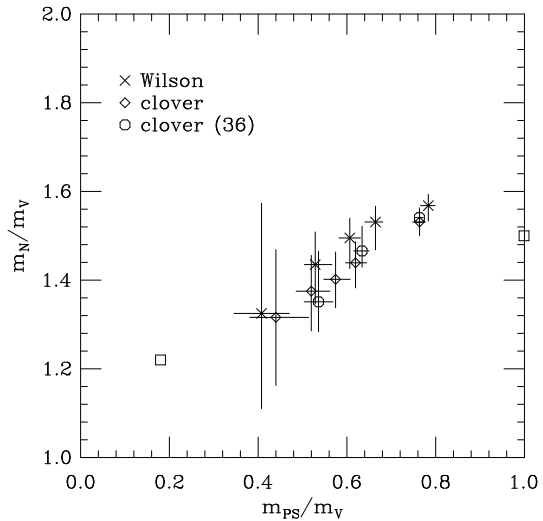

\centerline{
\ewxy{edin.ps}{70mm}
}
\caption{Edinburgh plot for both actions. The physical and heavy quark
points are marked by the left-hand and right-hand $\Box$ respectively.
\label{fig:edin_plot}}
\end{figure}

The Edinburgh plot for the data is shown in \Fig{fig:edin_plot}.  The
results for the two actions are broadly consistent, showing a trend
towards the physical value of $m_N/m_V$ with decreasing $m_{\PS}$.

\begin{figure}[hbt]
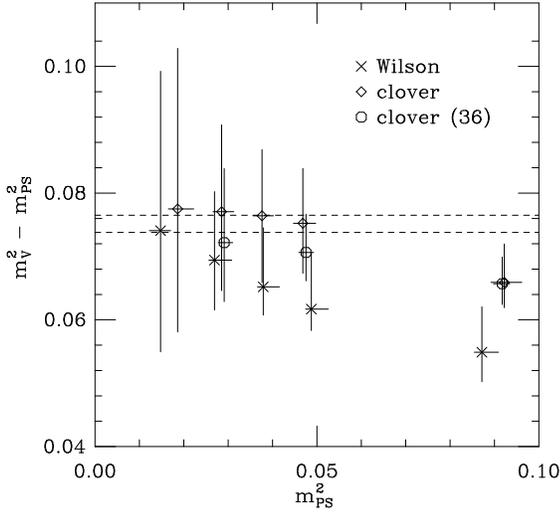

\centerline{
\ewxy{mrho2_minus_mpi2_vs_mpi2.ps}{70mm}
}
\caption{$m_V^2 - m_{\PS}^2$. The experimental range is indicated
by the horizontal lines.
\label{fig:meson_splitting}}
\end{figure}

For the comparison of the two actions, we have concentrated on
quantities which might be more sensitive to $O(a)$ effects, such as
mass splittings.   In \Fig{fig:meson_splitting}, we plot the difference
between the squares of the vector and pseudoscalar masses as a function
of the square of the pseudoscalar mass in lattice units.
Experimentally, this quantity is almost independent of the quark masses
for both heavy-light and light-light mesons.  We denote the range of
experimental values by the horizontal lines, setting the scale using
the string tension (\ie, $a^{-1} = 2.73\er{5}{5}$~GeV~\cite{hadrons_lett}).
In the chiral limit, the
splitting for the Wilson action is comparable to the experimental
values, while, at heavier quark mass, it is considerably smaller. For
the clover action, there is a much weaker dependence on the quark mass
and the values are roughly consistent with experiment.

\begin{figure}[htb]
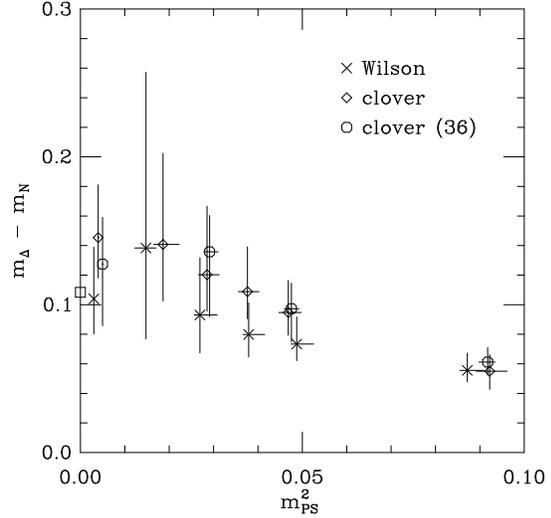

\centerline{
\ewxy{mdelta_minus_mn_vs_mpi2.ps}{70mm}
}
\caption{Baryon mass splitting.  The experimental value is marked with
a $\Box$.
\label{fig:baryon_splitting}}
\end{figure}

The differences between the $\Delta$ and nucleon masses are shown in
\Fig{fig:baryon_splitting}.  The leftmost points are obtained from the
chiral extrapolations of the individual masses.  These values, as well
as those from lighter quark masses are in good agreement with the
experimental value shown.

In \Tab{tab:scales}, we present various determinations of the lattice
scale.  The values obtained from hadron masses are from correlated
linear extrapolations to the chiral limit.  The scales obtained from
$m_\rho$ agree well with the scale from the string tension, whereas
those from the baryon masses, while self-consistent, are lower.
\begin{table}[ht]
\caption{Scales determined from various quantities.
\label{tab:scales}}
\begin{tabular}{lccc}
\hline
& & \multicolumn{1}{c}{\rule{0em}{1.0em}$a^{-1}$~GeV}      \\
\hline
Action & Wilson & \multicolumn{2}{c}{clover} \\
$N_{\rm cfgs}$ & 18 & 18 & 36 \\
\hline
$m_\rho$   & 2.77\err{9}{23}  & 2.57\err{16}{16} & 2.68\err{11}{12} \\
$m_N$      & 2.39\err{10}{12} & 2.45\err{18}{11} & 2.38\err{10}{15} \\
$m_\Delta$ & 2.48\err{8}{15}  & 2.33\err{10}{10} & 2.36\err{14}{13} \\
\hline
\multicolumn{2}{l}{string tension} & \multicolumn{1}{c}{2.73\err{5}{5}} \\
\hline
\end{tabular}
\end{table}


\pagebreak

\subsection{Decay Constants}

The results presented for the decay constants below are all obtained
using propagators with local sinks.

\subsubsection{Pseudoscalar Decay Constant}

This is defined through the matrix element of the axial current
\begin{equation}
f_{\PS} = Z_A f_{\PS}^L
      = \frac{Z_A \langle 0 | \: A_4 \: | \PS \rangle}{m_{\PS}}
\end{equation}
and our normalisation is such that the physical value is 132 MeV.  We
calculate $f_{\PS}^L$ using
\begin{equation}
f_{\PS}^L = \frac{1}{m_{\PS}}
\sqrt{
  \frac{\langle A_4 \: A_4 \rangle}{\langle P \: P \rangle}
  {\left| \langle 0 | \: P \: | \PS \rangle \right|}^2
}
\end{equation}
with \mbox{$A_\mu = \bar{q} \gamma_\mu \gamma_5 q$} and
\mbox{$P = \bar{q} \gamma_5 q$}.  We fix $m_{\PS}$ and
$\langle 0 | \: P \: | \PS \rangle$ from the
pseudoscalar propagator.
In perturbation theory, the lattice renormalisation factors
are $Z_A^W \simeq 1 - 0.13 g^2$  and $Z_A^C \simeq 1 - 0.02
g^2$~\cite{pittori}. Using the bare coupling gives $Z_A^W \simeq 0.87$
and $Z_A^C \simeq 0.98$, whereas, if we use the  ``effective coupling''
proposed by Lepage and Mackenzie~\cite{lm}, we get $Z_A^W \simeq 0.78$
and $Z_A^C \simeq 0.97$.

We have investigated the dimensionless quantity $f_{\PS} / m_V$, and
find it varies only slowly with quark
mass, in agreement with the experimental observation that
$f_\pi/m_\rho$ (0.17) is approximately the same as
$f_K/m_{K^\ast}$ (0.18).  In \Tab{tab:fPS_to_mV}, we show the values
obtained from linear extrapolations to the chiral limit.
For both actions, the chiral extrapolations are roughly consistent with
experiment. The fact that the data from the clover action has larger
statistical errors is compensated by the smaller uncertainty in the
lattice renormalisation factor, $Z_A^W$.
This means that we can, in principle, obtain more accurate predictions
using the clover action by increasing statistics.

\begin{table}[bht]
\caption{Chiral extrapolations of $f_{\PS}/m_V$.
\label{tab:fPS_to_mV}
}
\begin{tabular}{lcccc}
\hline
Action & Wilson & \multicolumn{2}{c}{clover} \\
$N_{\rm cfgs}$ & 18 & 18 & 36 \\
\hline
\rule[-0.55em]{0em}{1.65em}$f^L_{\PS} / m_V$
	& 0.21\er{2}{3} & 0.13\er{4}{3} & 0.15\er{2}{2} \\
\rule[-0.55em]{0em}{1.65em}$Z^{\rm bare}_A f^L_{\PS} / m_V$
	& 0.18\er{2}{3} & 0.12\er{4}{2} & 0.15\er{2}{2} \\
\rule[-0.55em]{0em}{1.65em}$Z^{\rm eff}_A f^L_{\PS} / m_V$
	& 0.16\er{2}{3} & 0.12\er{4}{2} & 0.15\er{2}{2} \\
\hline
\end{tabular}
\end{table}

\subsubsection{Vector Decay Constant}
\raggedbottom
We define $f_V$ through
\begin{equation}
\frac{1}{f_V} \, \epsilon_i = \frac{Z_V}{f_V^L} \, \epsilon_i
   = \frac{Z_V}{m_V^2} \langle 0 | V_i | V \rangle
\end{equation}
with \mbox{$V_i = \bar{q}\gamma_i q$}.
\Tab{tab:fV} shows the lattice results for both actions from 18
configurations.  The values depend only weakly on quark
mass for both actions.
The results obtained from a linear extrapolation to the  chiral limit
are given in the final row. Using the bare~(``effective'') coupling,
these correspond to  $1/f_V^W = 0.39\er{2}{1} \; (0.33\er{1}{1})$ and
$1/f_V^C = 0.34\er{2}{1} \; (0.31\er{2}{1})$.  These are similar
to the experimental value, $1/f_\rho = 0.28(1)$.  Using the conserved
vector current would remove the uncertainty in the normalisation
for the Wilson action.

\begin{table}[hbt]
\caption{$1/f_V$ as a function of $\kappa$.
\label{tab:fV}
}
\begin{tabular}{cccc}
\hline
\multicolumn{2}{c}{Wilson} & \multicolumn{2}{c}{clover} \\ \hline
\rule[-0.55em]{0em}{1.65em}$\kappa$& $1/f_V^L$ & $\kappa$ & $1/f_V^L$ \\ \hline
0.15100 & 0.40\er{1}{1} & 0.14144 & 0.33\er{1}{2} \\
0.15200	& 0.43\er{1}{1} & 0.14226 & 0.36\er{2}{1} \\
0.15230	& 0.44\er{1}{1} & 0.14244 & 0.36\er{2}{1} \\
0.15260	& 0.45\er{1}{1} & 0.14262 & 0.37\er{2}{1} \\
0.15290	& 0.47\er{2}{1} & 0.14280 & 0.36\er{2}{1} \\ \hline
$\kappacrit^W$ & 0.47\er{2}{1} & $\kappacrit^C$ & 0.38\er{3}{1} \\ \hline
\end{tabular}
\end{table}

\pagebreak



\section{CONCLUSIONS}

For both actions, mass splittings and decay constants typically agree
with experiment within errors.  We find broadly consistent scales from
different quantities, though the baryon scales appear somewhat low.

We have shown that the clover action can be used to study light
hadrons.  The computational overhead is small, except that the
increased statistical fluctuations mean that we require smearing or
more configurations for comparable errors.  We see a few hints of
improvement, though, in general, the results are similar to those from
the standard Wilson action.  This suggests that, at \mbox{$\beta =
6.2$}, $O(a)$ effects may be small for light hadrons.

\subsection*{Acknowledgements}

This research is supported by the University of Edinburgh, by Meiko
Limited and by the UK Science and Engineering Research Council under
grants GR/G~32779, GR/H~49191 \& GR/H~01069. The author also
acknowledges the personal support of SERC through the award of a
Postdoctoral Fellowship.




\end{document}